\documentclass[10pt]{article}
\usepackage{ametsoc2col}
\newcommand{\myabstract}{The Facebook Page Network (FPN) is a platform for Businesses, Public Figures and Organizations (BPOs) to connect with individuals and other BPOs in the digital space. For over a decade scale-free networks have most appropriately described a variety of seemingly disparate physical, biological and social real-world systems unified by similar network properties such as scale-invariance, growth via a preferential attachment mechanism, and a power law degree distribution  $P(k) \propto ck^{-\lambda}$ where typically  $2<\lambda<3$. In this paper we show that both the Facebook Page Network and its BPO-BPO subnetwork suggest power law and scale-free characteristics. We argue that social media analysts must consider the logarithmic and non-linear properties of social media audiences of scale.}
\begin{document}
\title{\textbf{\large{On the Topology of the Facebook Page Network}}}
\author{\textsc{R.E. Slattery,}
				\thanks{\textit{Corresponding author address:} 
				Enovai, Inc.
				Hartford, CT, USA 
				\newline{E-mail: rslattery@enovai.com}}\quad\textsc{R.R. McHardy and R. Bairathi}\\
\textit{\footnotesize{Enovai, Inc., Hartford, Connecticut;  Emergence API}}
}
\ifthenelse{\boolean{dc}}
{
\twocolumn[
\begin{@twocolumnfalse}
\amstitle
\begin{center}
\begin{minipage}{13.0cm}
\begin{abstract}
	\myabstract
	\newline
	\begin{center}
		\rule{38mm}{0.2mm}
	\end{center}
\end{abstract}
\end{minipage}
\end{center}
\end{@twocolumnfalse}
]
}
{
\amstitle
\begin{abstract}
\myabstract
\end{abstract}
\newpage
}
\section{Introduction}
Social media's proliferation has led to a trove of novel analytic and topological data for eager network scientists and has provided an interested public a real-world context for network theory principles. Today social platforms encourage Brands, Public Figures and Organizations (BPOs) to participate alongside individuals in the social space to deepen on- and offline relationships. The resulting increase of digital attention and interaction has led BPOs to invest more heavily into social outreach; the most successful BPOs over time have best aligned social platforms with more traditional resources and objectives. Efficient management of digital public relations processes has become critical to the day-to-day operations of many BPOs. 

The Facebook Page Network (FPN) in particular is one of the primary avenues where BPOs engage and, along with Twitter, YouTube, and LinkedIn (among others), has fundamentally changed the mechanisms of BPO online presence management.  The FPN provides a space where individuals may interact with various pieces of digital media, comment on publications by the BPO, and share certain items with friends within their own personal network. If an individual would like to receive updates or display an affiliation or affinity toward a particular BPO, one may do so by \textsc{Liking} the BPO, which is then publicly displayed on the individual's own page. Individuals may choose to interact with a BPO digitally for a variety of reasons: to genuinely support the BPO, access locked content, receive member benefits, derive entertainment value, acquire information, or signal things about themselves such as interests and associations. BPOs may choose to publicly interact with other BPOs for a variety of reasons as well: to reflect business associations, affiliate with a particular cause, or lure a potential response from the target BPO. Like some individuals BPOs may use this opportunity to enhance their public online appearance.  

Networks are represented as graphs by an interaction mapping function between a collection of nodes and links. In the case of the Facebook Page Network, a node is an entity, either an individual or a BPO, and a link is a directed \textsc{Like} between them. The total number of \textsc{Likes} is the degree of the node, and the degree distribution can be obtained by calculating the frequency of these degrees. Classical network models of the mid-20th century such as the Erdos-Renyi Random Graphs (ERRGs) distribute degrees normally:  an average node has an average number of links and contains little discernible topological structure [1]. While such a model may be appropriate for highly entropic and structurally simple network topologies, real-world self-organizing systems have more complex behavior and lead to a variety of topological non-trivialities. Experiments by Milgrim in the 1960s [2] and Watts and Strogatz in the 1990s [3] show that social networks in particular often have small-world character: tremendously large networks can be traversed in a short number of steps, nodes with mutual links cluster together, connections are formed more heterogeneously, and sometimes nodes even synchronize. Public consciousness is well aware of the six degrees of separation phenomenon [4], and the recent pervasiveness of social media has undoubtedly exacerbated a feeling of increased global connectedness. 

For over a decade scale-free networks have most appropriately described a variety of these seemingly disparate physical, biological and social real-world complex systems unified by similar network properties such as a power law degree distribution, growth via a preferential attachment mechanism, and scale-invariance. Scale-free networks were independently recognized in 1999 by the Faloustos brothers in the mapping of the Internet [5], and more famously by the Barabasi workgroup at Notre Dame in the mapping of the World Wide Web [6]. These networks are classified by a highly heterogeneous degree distribution best described by a power law  of the form $P(k) \propto ck^{-\lambda}$, where $\lambda$ is typically found $2<\lambda<3$ when plotted to logarithmic scale. Scale-free networks are less structurally entropic than their ERRG counterparts due to a characteristic topological property: the power law distribution produces few highly connected ‘hubs’ and many sparsely connected nodes, while maintaining this heterogeneity across orders of magnitude. Scale-free topologies produce ultra small-world networks with an average path length of $log⁡(N)$, which makes these networks readily navigable at very large sizes [7]. Scale-free networks are fragilely robust to random failure, but are acutely vulnerable to targeted attack and have an effective epidemic threshold of zero. Scale-free networks have been suggested in the scientific collaboration and citation networks [8], the distribution of United States airports [9],  the potential energy landscape of folding proteins [10], the protein-protein interaction network [11], the metabolic pathway network [12], the spread of sexual diseases [13], and a variety of linguistic, neuronal, social, biological and technical networks.

Power law heterogeneity, such as Pareto's 80/20 law [14], the Yule-Simon distribution [15-16], Zipf's distribution [17], de Solla Price's cumulative advantage [18], and Merton's Matthew Effect [19] all involve a process in which the rich get richer. Canonical scale-free networks obtain their topology by an analogous mechanism, Barabasi-Albert Preferential Attachment (BAPA), an algorithm that rewards more connected nodes in linear relation based on two assumptions: (1) the network is growing; (2) nodes receive new links with a linearity relative to their degree
	\begin{equation} \label{eq:1}
p_i = \frac{k_i}{\sum_{j}{k_j}}
\end{equation}
where $k$ is the degree of a particular node. This process rewards the most connected nodes, reinforcing the network's characteristic heterogeneity, and reveals a power law with an exponent  of $2<\lambda<3.$ In the FPN and BPO-BPO networks this would suggest that the most \textsc{Liked} BPOs have the highest probability of obtaining new \textsc{Likes} as the network grows. Beyond the canonical BAPA mechanism, scale-free networks can be induced by a variety of more boundedly rational processes: random walks [20]; optimization [21]; cooperation [22]; competition [23], among others. Power laws, scale-free networks and BAPA are not without detractors [24] and are notoriously polemical, but the basic phenomenon has held up despite tweaks to the canonical form. In this paper we demonstrate that both the FPN and BPO-BPO networks suggest a power law distribution and display characteristics consistent with scale-free networks.

\section{Method}

Facebook and a host of other social networks offer rich Application Programming Interfaces (APIs) to provide researchers, developers and entrepreneurs an opportunity to access, discover and implement features of social databanks. We obtained the total \textsc{Likes} for each BPO via the Facebook Developer API and aggregated to obtain the network-wide distribution. In the specific case where a BPO \textsc{Likes} another BPO, this \textsc{Like} is publicly listed on the former BPO's page. The BPO-BPO network is orders of magnitude smaller than the FPN and represents a hub-like core. We traverse this network using a breath-first search strategy to reconstruct the network topologically.  Preliminary findings obtained in August 2012 on a $\sim$ 400,000 node network suggested that both the FPN and BPO-BPO networks displayed power-law properties, and the BPO-BPO network in particular within an exponent in the range characteristic of scale-free networks. In April  2013 we obtained a far more complete network sample containing $\sim 4.8$ million nodes and  $\sim 24$ million links and confirmed our earlier results. 

\section{Results}

We found that the FPN and the BPO-BPO hub subnetwork both to suggest power law scaling and scale-free characteristics, reported in $\mathbf{Fig.}$ $\mathbf{1}$ and $\mathbf{Table}$  $\mathbf{1}$. We obtained the degree distribution of the FPN across  $\sim 4.8$ million nodes and  $\sim 24$ million links and found the linear range of the FPN distribution obeys power law where $\lambda = 1.53$ with an  ${R}^2 = 0.913$, slightly sharper than a characteristic scale-free network where $2<\lambda<3.$ The distribution is initially stagnant before reaching a critical threshold, in this case $\sim 10^2$. 

We reconstructed the BPO-BPO hub subnetwork topologically and found that the distribution revealed a power law where $\lambda = 2.25$ along the linear range with an  ${R}^2 = 0.995$, consistent with scale-free networks and suggestive of linear BAPA. 

\section{Discussion}

\begin{figure*}[t]

  \noindent\includegraphics[width=44.2pc,angle=0]{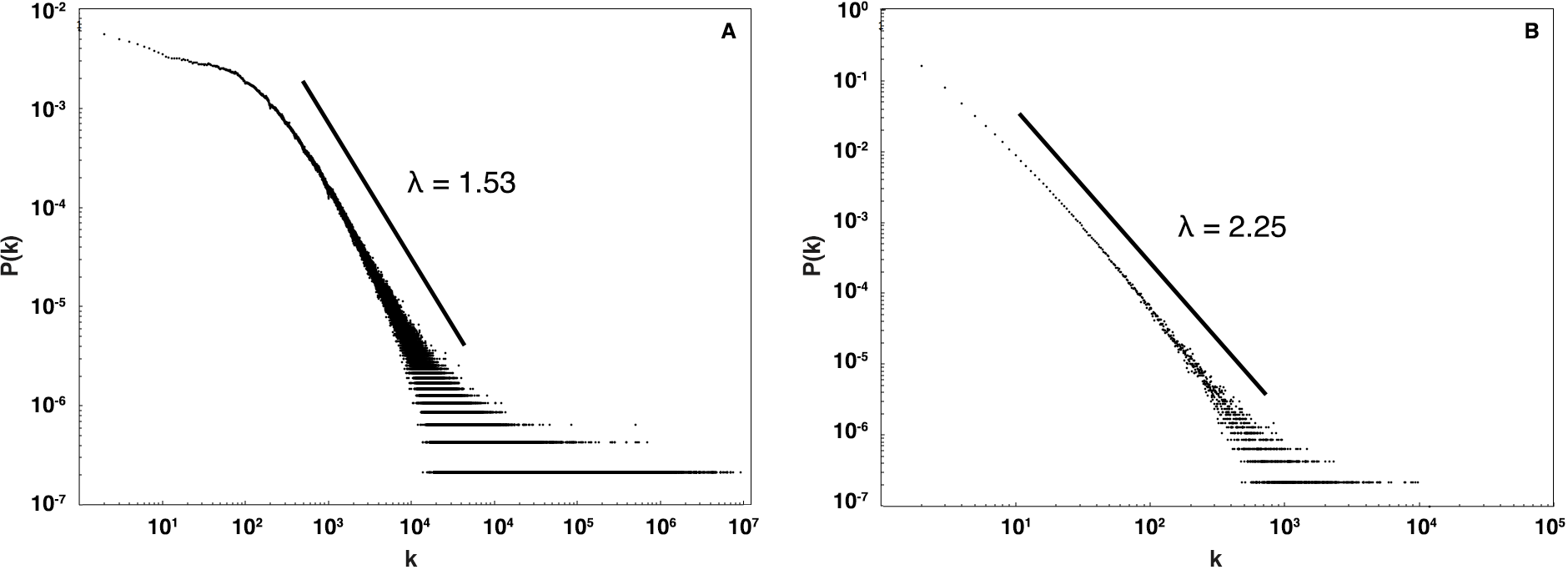}\\
  \caption{$\mathbf{A.}$ The distribution of the Facebook Page Network (FPN) suggests a power law $P(k) \propto ck^{-\lambda}$ at logarithmic scale with an exponent $\lambda = 1.53$ after BPOs overcome a threshold $\sim 10^2$. The x-axis $k$ denotes the degree of the node, in this case the number of \textsc{Likes} of a BPO. The y-axis $P(k)$ represents the probability of this degree occurring within the network. $\mathbf{B.} $ The distribution of the BPO-BPO Network within the FPN suggests a power law with an exponent  $\lambda = 2.25$ along the linear range when re-constructed topologically, characteristic of a scale-free network. }\label{f1}
\end{figure*}

\begin{table}[t]
\caption{Coefficients from the Facebook Page Network (FPN) and the Brands, Public Figures and Organizations BPO-BPO  hub-like core following the form $P(k) \propto ck^{-\lambda}.$}\label{t1}
\begin{center}
\begin{tabular}{ccccrrcrc}
\hline\hline
$network$ &  $c$ & $\lambda$ & $R^2$\\
\hline
FPN  & 2E+07 & 1.531& 0.913 \\
BPO-BPO  & 9E+06 & 2.252 & 0.995\\
\hline
\end{tabular}
\end{center}
\end{table}
Both the FPN and BPO-BPO networks exhibit logarithmically heterogeneous power law scaling, and suggest a complex topology with scale-free character. In both cases the goodness-of-fit was over 90\% along the linear range. The topological structure of a network can influence opinion formation, communication, availability, access and exposure to certain digital items. Unlike many physical and biological networks, social network dynamics often concern conscious decision-making processes between agents. Both the FPN and BPO-BPO network coalesce around a minority of hubs, which may restrict information exchange and make it more difficult to acquire through less connected channels. The potential reach of a BPO is also subject to topological constraints: the most connected BPOs have the highest probability of obtaining new \textsc{Likes}. More connected BPOs have greater interaction potential with more visible content, even if the content is engagingly identical when compared to a less-connected BPO. 

Future social analytic researchers should be considerate of these logarithmic and non-linear properties when appropriately analyzing social networks. For example, the commonly cited Engagement Rate (ER) of a particular piece of media is calculated as

	\begin{equation} \label{eq:1}
ER = \frac{Interactions}{Likes_{BPO}}
\end{equation}\*which fails to consider the logarithmic and non-linear properties of both BPO reach and engagement dynamics. Implicit in the canonical ER equation is the assumption that both interaction and reach data are normally distributed. We show this is not the case, and the power law distribution renders cross-sectional analysis as arbitrary. In this particular instance, logarithmic econometric regression analysis could provide a more appropriate benchmarking measure between BPOs.

\section{Conclusion}
We present a network analysis of the Facebook Page Network and the BPO-BPO hub-like core. In both cases we found that the networks suggest a power law distribution, and in particular the BPO-BPO network has scale-free character when topologically reconstructed. The Facebook Page Network and BPO-BPO subnetwork are subject to the effects of scale-free topologies: heterogeneous in content distribution, preferential in growth, and leaning toward further hub dominance. We suggest that social media analysts must consider logarithmic and non-linear properties of social media audiences of scale.

\begin{acknowledgment}
Produced with the Emergence API, for more information please visit http://www.emergenceapi.com. In memory of A. Syzchowski. 

Corresponding author: rslattery@enovai.com.
\end{acknowledgment}

\ifthenelse{\boolean{dc}}
{}
{\clearpage}

\ifthenelse{\boolean{dc}}
{}
{\clearpage}
\bibliographystyle{ametsoc}
\bibliography{references}

\clearpage
\centerline{REFERENCES}
\vspace{5 mm}

\textbf{[1] } Erdos, P. and A. Renyi. Publ. Meth. Inst. Hungar. Acad. Sci. 5:17-61 (1960). 

 \textbf{[2] } Travers, J. and S. Milgrim. Sociometry, Vol. 32, No. 4 (1969).

\textbf{[3] } Watts, D.J. and S.H. Strogatz. Nature  393, 440 (1998).

\textbf{[4] } Karinthy, F. (1928), Guare, J. (1990).

\textbf{[5] } Faloustos, M., P. Faloustos and C. Faloustos. ACC SIGCOMM, 1999.

\textbf{[6] } Albert, R., H. Jeong and A.-L. Barabasi, Nature, 401 (1999).

 \textbf{[7] } Cohen, R. and S. Havlin. Phys. Rev. Lett. 90.058701 (2003). 

\textbf{[8] } Newman M. Phys. Rev. E 64, 016131 (2001).

\textbf{[9] } Guimera, R. PNAS 102.22 (2005). 

\textbf{[10] } Doyle, J. Phys. Rev. Lett. 88 (2002).

\textbf{[11] } Uetz, P. et al. Nature 403 (2003). 

\textbf{[12] } Jeong. H, S. Mason, A.-L. Barabasi, and Z. Oltvati.  Nature 411 (2001). 

\textbf{[13] } Liljeros, F. et al. Nature Vol. 411, (2001).

\textbf{[14] } Pareto, V. (1896) via Juran, J. (1940).

\textbf{[15] } Yule, G. Phil. Trns. Royal Soc. Ser. B (1925).

\textbf{[16] } Simon, H. Biometrika 42 (1955).

\textbf{[17] } Zipf, G. Hrvrd. Press. (1932).

 \textbf{[18] } de Solla Price, D. Joural. Americ. Soc. Inform. Sci. 27 (1976). 

\textbf{[19] } Merton, R. Science, 159 (1968).

\textbf{[20] } Vazquez, A. Physical Review E 67.5 (2003).

\textbf{[21] } Berger, N., C. Borgs, J.T. Chayes, R.M. D'Souza, and R.D. Kleineberg. PNAS 104.15 (2007).

\textbf{[22] } Santos, F.C. and J.M. Pacheco. Phys. Rev. Letters 95.5 (2005).

\textbf{[23] } Berger, N., C. Borgs, J.T. Chayes, R.M. D'Souza, and R.D. Kleineberg. Comb. Prob. and Computing 15. 5-6 (2005).

\textbf{[24] } Stumpf, M. P., Wiuf, C., and May, R. M. PNAS USA, 102(12) (2005).

\end{document}